\documentclass[aps,prd,twocolumn,superscriptaddress,floatfix]{revtex4-2}
\usepackage{graphicx}
\usepackage{makecell} 
\usepackage{booktabs} 
\usepackage{braket}
\usepackage{amsfonts} 
\usepackage[margin=1in]{geometry}
\usepackage{tikz}
\usepackage{quantikz}

\usepackage{graphicx}
\usepackage{amstext} 
\usepackage{amsmath} 
\usepackage{array}   
\newcolumntype{L}{>{$}l<{$}} 
\usepackage{pstricks}
\usepackage{color}
\usepackage{placeins}
\usepackage{hyperref}

\begin{document}

\title{Search for \(t\bar tt\bar tW\) Production at \(\sqrt{s}=13\;\mathrm{TeV}\) Using a Modified Graph Neural Network at the LHC}

\author{Syed Haider Ali}
\email[]{syedhaider.ali2021453@gmail.com}
\affiliation{Department of Physics \& Applied Mathematics , Pakistan Institute of Engineering and Applied Sciences (PIEAS) , P. O. Nilore , 45650 , Islamabad, Pakistan.}
\affiliation{Exp. High Energy Physics Department (EHEPD) , National Centre for Physics (NCP), Shahdra Valley Road, 44000, Islamabad, Pakistan. }
\author{Ashfaq Ahmad}
\email[]{Ashfaq.Ahmad@cern.ch}
\affiliation{Exp. High Energy Physics Department (EHEPD) , National Centre for Physics (NCP), Shahdra Valley Road, 44000, Islamabad, Pakistan. }
\author{Muhammad Saiel}
\affiliation{Exp. High Energy Physics Department (EHEPD) , National Centre for Physics (NCP), Shahdra Valley Road, 44000, Islamabad, Pakistan. }
\author{Nadeem Shaukat}
\affiliation{Department of Physics \& Applied Mathematics , Pakistan Institute of Engineering and Applied Sciences (PIEAS) , P. O. Nilore , 45650 , Islamabad, Pakistan.}



\begin{abstract}
The simultaneous production of four top quarks in association with a $W$ boson at $\sqrt{s} = 13$  TeV is an rare SM process with a next-to-leading-order (NLO) cross-section of $(6.6^{+2.4}_{-2.6} {ab})$\cite{saiel}. Identifying this process in the fully hadronic decay channel is particularly challenging due to overwhelming backgrounds from $t\bar{t}, t\bar{t}W, t\bar{t}Z$, and triple-top production processes. This study introduces a  modified physics informed Neural Network, a hybrid graph neural network (GNN) enhancing event classification. The proposed model integrates Graph layers for particle-level features, a custom Multi Layer Perceptron(MLP) based global stream with a quantum circuit and cross-attention fusion to combine local and global representations. Physics-informed Loss function enforce jet multiplicity constraints, derived from event decay dynamics. Benchmarked against conventional methods, the GNN achieves a signal significance $(S/\sqrt{S+B})$ of $0.174$ and ROC-AUC of 0.974, surpassing BDT's significance of  $0.148$ and ROC of $0.913$, while XGBoost achieves a significance of $0.149$ and ROC of $0.920$. The classification models are trained on Monte Carlo (MC) simulations, with events normalized using cross-section-based reweighting to reflect their expected contributions in a dataset corresponding to $350\;$fb$^{-1}$  of integrated luminosity. This enhanced approach offers a framework for precision event selection at the LHC, leveraging high dimensional statistical learning and physics informed inference to tackle fundamental HEP challenges, aligning with ML developments.  
 
\end{abstract}


\maketitle

\section{Introduction}
The top quark was discovered in top anti-top pair production at Tevatron at Fermilab ~\cite{D0:1995jca,Abe:1995hr}, which has the highest mass $172.52\;\pm\;0.14\;(stat)\pm\;0.30\;(syst)$ GeV\cite{cmstop2024combination} among the particles in the Standard Model. The production of single top quark occurs through the electroweak interaction mediated by the $tWb$ vertex and was also discovered at the Tevatron ~\cite{Aaltonen:2009jj,Abazov:2009ii}. The present level of experimental sensitivity has now advanced from early evidence to the full observation of the Standard Model four top quark production process. In the Standard Model, the production of four top quarks ($t\bar{t}t\bar{t}$) is among the rarest processes occuring in proton-proton collisions currently accessible at hadron colliders. The Standard Model cross section has been determined at next-to-leading order (NLO) in both QCD and electroweak theory, incorporating soft‐gluon emission effects at next‐to‐leading logarithmic precision, to be $13.4\;^{+1.0}_{-1.8}$ fb at $\sqrt{s} = 13$ TeV\cite{vanbeekveld2025,CMS:2023fourtop} .\\
Despite its rarity, the CMS experiment has reported evidence for and, more recently, the observation of this channel using Run 2 data at the LHC. The CMS Collaboration first presented evidence for four-top production in leptonic and semi-leptonic final states, achieving a significance of 3.9 standard deviations~\cite{CMS:2019udw}. Most recently, CMS has reported the first observation of the $t\bar{t}t\bar{t}$ production process at 13 TeV with an observed significance of 5.6 standard deviations~\cite{CMS:2023fourtop}. These measurements validate the SM prediction and open a sensitive window to potential new physics in multi-top interactions. Moreover, there are also several processes involving four top quark production in association with other particles, which have even lower cross sections and remain experimentally elusive. Rare event classification plays a crucial role in high-energy physics by enabling the detection of processes that could provide indirect evidence of physics beyond the Standard Model. These rare processes often occur at extremely low cross-sections, making their identification amidst overwhelming Standard Model backgrounds a significant challenge\cite{Leonardo2016Search}. One such rare process is the production of four top quarks  in association with a $W$ boson ($t\bar{t}t\bar{t}W$) at $\sqrt{s} = 13$ TeV, a Standard Model process with an NLO cross-section of $6.6^{+2.4}_{-2.6}ab$\cite{saiel}. Because the associated production channel \(pp \to t\bar{t}t\bar{t}W\) samples even higher partonic center-of-mass energies and allows for multiple insertions of contact operators, it is uniquely sensitive to four-fermion interactions of the form \((\bar{q}q)(\bar{t}t)\). In particular, deviations in the rate or kinematic distributions of \(t\bar{t}t\bar{t}W\) can constrain the Wilson coefficients of these dimension-six operators more stringently than conventional \(t\bar{t}\) measurements, thanks to the enhanced energy growth and interference effects highlighted in~\cite{zhang2018constraining}. This makes the fully hadronic \(t\bar{t}t\bar{t}W\) channel an especially powerful probe of anomalous top-quark couplings in the SMEFT\cite{Brivio_2019_SMEFT, Grzadkowski_2010SMEFT} framework. However, identifying it in the fully hadronic decay mode is particularly challenging due to significant background contamination from dominant SM processes like $t\bar{t}$ with SM cross-section order of magnitude higher than the simultaneous production of four top quarks in association with a $W$ boson \cite{saiel}.\\
\\
Event classification techniques in high-energy physics have traditionally relied on cut-based selections and multivariate methods such as Boosted Decision Trees (BDTs) \cite{boostingseminal} and XGBoost\cite{Xgboost}. These methods have been widely used due to their interpretability and computational efficiency. However, while effective for some tasks, these approaches struggle to model the intricate correlations between multiple final-state particles, particularly in high-dimensional datasets\cite{BDTLimitations}. High-energy collision events involve a complex interplay between jet kinematics, missing transverse energy, and multi-jet angular separations. A more sophisticated approach is needed to fully capture these dependencies and optimize signal-background discrimination.\\
\\
Deep learning has revolutionized event classification in HEP by enabling models to extract hidden patterns from large datasets without relying on handcrafted features. Neural networks have demonstrated superior performance in jet tagging, event selection, and anomaly detection at the LHC, offering a powerful alternative to traditional machine learning methods. Transformers\cite{vaswani2023attentionneed}, in particular, have emerged as a promising architecture for analyzing high-dimensional physics data. Unlike conventional convolutional\cite{CNN_Seminal} or recurrent\cite{RNN_Seminal} networks, transformers leverage self-attention mechanisms to dynamically weigh input features based on their importance, making them highly effective for learning inter-jet correlations\cite{Builtjes2022Attention}. These properties have led to their successful application in jet energy reconstruction, top-tagging, and particle trajectory prediction\cite{Zhang2024Reconstruction}.\\
\\
Deep learning models, despite their remarkable success in high-energy physics classification tasks, inherently lack built-in physics constraints. This limitation can lead to outputs that violate fundamental principles such as energy conservation and angular momentum invariance, which are critical in ensuring the physical reliability of predictions. To address this issue, \textbf{Physics-Informed Neural Networks (PINNs)} have been introduced as a framework that integrates domain-specific physics constraints directly into the model’s loss function\cite{PINNs_Seminal}. By embedding conservation laws and differential equations into the learning process, PINNs ensure that the learned representations remain not only physically consistent but also improve generalization to unseen data\cite{Farea2024Understanding}. These models have shown great success in maintaining the stability of constrained Hamiltonian systems and preserving energy conservation principles\cite{Kaltsas2024Constrained}. The synergy between \textbf{transformers and PINNs} presents a compelling approach to rare event classification, as it allows deep learning models to benefit from both the self-attention mechanism’s ability to capture long-range dependencies and the robustness of physics-driven regularization, making them more interpretable and effective for high-dimensional physics datasets.\\
\\
This paper presents a modified GNN for rare event classification in HEP. The proposed model integrates GINEConv\cite{hu2020strategiespretraininggraphneural, xu2019powerfulgraphneuralnetworks} layers for particle-level features, a quantum circuit with angle encoding and entanglement for global features\cite{schuld2019quantum}, and cross-attention fusion, with physics-informed Loss enforcing  jet multiplicity constraints. The model is trained on Monte Carlo simulated data, with cross-section-based reweighting applied to reflect realistic event yields corresponding to $350$ fb$^{-1}$ of integrated luminosity\cite{saiel}. By benchmarking the GNN against BDTs, XGBoost, and conventional transformers, we demonstrate its superior classification performance in distinguishing the signal in it's fully hadronic decay mode for optimizing $t\bar{t}t\bar{t}W^-$ searches at the LHC.\\
\\
The rest of this paper is structured as follows. The next section details the dataset, feature selection, and event simulation process. This is followed by an in-depth discussion of the proposed GNN architecture, including its quantum and classical components and physics-informed constraints. The results compare the model’s performance against conventional methods, highlighting improvements in classification significance. Finally, we examine the internal behavior of the architecture to better understand how its different components contribute to the final decision-making process. This work represents a step forward in applying AI-driven methodologies to rare event classification for improved sensitivity in BSM searches at the Large Hadron Collider and future high-luminosity experiments.

\section{Dataset and Event Simulation}\label{sec2}

\subsection{Monte Carlo Simulation Setup}

The simulated event samples of signal and backgrounds are generated using MC events generators. The $t\bar{t}t\bar{t}W^-$ signal is generated at LO using MadGraph5 aMC@NLO v2.9.15 \cite{madgraphNLO}. The NLO calculation of $6.6^{+2.4}_{-2.6}$ ab was used to normalize the simulation \cite{saiel}. The quoted uncertainty includes the variation of factorization and renormalization scales used in the calculation of matrix elements (ME), as well as the dependence on the choice of parton distribution functions (PDFs) \cite{saiel42}. The signal events are decayed into final state particles using the mad spin card to include spin correlation \cite{saiel42}. Background process such as diboson ($WW$ and $WZ$), $t\bar{t}$, $t\bar{t}$ with a gauge boson, $t\bar{t}$ with Higgs and triple top quarks associated production ($t\bar{t}t\bar{q}$, $t\bar{t}tW$, $t\bar{t}t\bar{q}$, $t\bar{t}\bar{t}W$, $t\bar{t}\bar{t}\bar{q}$) are simulated using the same MadGraph5 version at LO \cite{saiel}. The table \ref{tab:cross_sections} shows the different processes and their corresponding MC generators along the cross section. For all process, NNPDF30\_nlo\_as\_0118 PDFs are used in calculation of matrix elements \cite{saiel44}. Hadronization and parton showering are simulated using Pythia8\cite{pythia8}. A fast simulation of the CMS detector, based on the Delphes\cite{delphes}, is used to process the simulated events.

\renewcommand{\arraystretch}{1.3} 

\begin{table}[h]
    \centering
    \caption{Summary of the signal and background process along with event generators and their corresponding production cross-section\cite{saiel}}
    \label{tab:cross_sections}
    \begin{tabular}{|c|c|c|}
        \hline
        \textbf{Process} & \textbf{Event Generator} & \textbf{Cross Section (pb)} \\
        \hline
        \textit{$t\bar{t}t\bar{t}W$}& MadGraph5+Pythia8 & $6.628\times10^{-6}$ (NLO)\\
        \hline
        $t\bar{t}$& MadGraph5+Pythia8 & $456$ (LO) \\
        \hline
        $t\bar{t}Z$& MadGraph5+Pythia8 & $0.5343$ (LO) \\
        \hline
        $WZ$ & MadGraph5+Pythia8 & $10.56$ (LO) \\
        \hline
        $t\bar{t}H$& MadGraph5+Pythia8 & $0.3678$ (LO) \\
        \hline
        $t\bar{t}W$& MadGraph5+Pythia8 & $0.1138$ (LO) \\
        \hline
        \textit{tripletop} & MadGraph5+Pythia8 & $6.932\times10^{-5}$ (LO)\\
        \hline
    \end{tabular}
\end{table}

\subsection{Event Selection and Feature Engineering}

\begin{figure}
    \centering
    \includegraphics[width=1\linewidth]{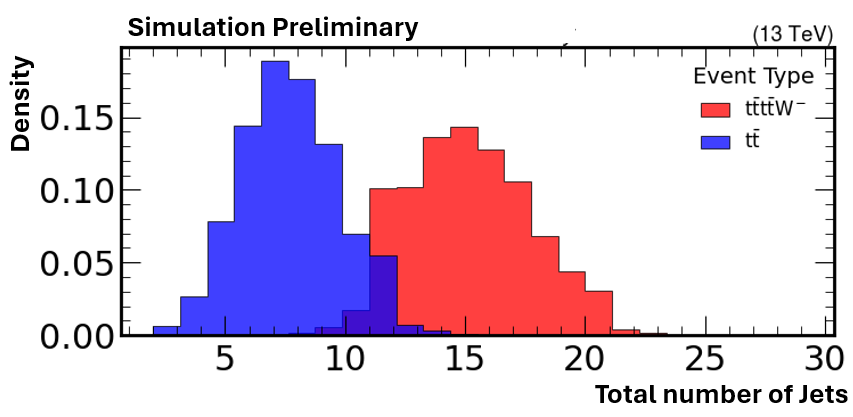}
    \caption{Jet Multiplicities of $t\bar{t}t\bar{t}W^-$ compared with the most abundant background ($t\bar{t}$) at $\sqrt{s} = 13$ TeV at CMS}
    \label{fig:jet_mult}
\end{figure}

This study considers the fully hadronic decay channel due to its higher branching ratio, despite the overwhelming background from ($t\bar{t}$), ($t\bar{t}W$), ($t\bar{t}Z$), and triple top quarks and its associated processes. Each of the b quarks form b-jets after hadronizing, while the light quarks form light jets, resulting in a final state where each top quark yields one b-jet and two light jets. Thus for the four top quark process we typically expect 4 b-jets and 10 light jets (8 from the hadronic decays of the W bosons from the top quark and 2 jets from the decay of the additional W). In contrast, the dominant background, $t\bar{t}$, typically produces fewer jets per event (for example, 2 b-jets and 4 light jets). These differences in decay dynamics are critical; the signal is characterized by a highly “jetty” environment, as depicted in  Figure \ref{fig:jet_mult}, that shows the jet multiplicity of the signal ($t\bar{t}t\bar{t}W^-$) with the most abundant background $t\bar{t}$. By focusing on the fully hadronic channel, this study aims to optimize the discrimination between signal and background to enhance the sensitivity of rare event searches.

The features described in table \ref{tab:dataset-variables} were determined suitable for the training of BDT based on the Exploratory Data Analysis (EDA) of the dataset as described in the upcoming section. Subsequently, since this study aims to establish a comparative analysis in multivariate analysis using traditional cut-based ML techniques and modern deep learning, the same features were used in training the Deep Learning(DL) architectures. For the BDT, trained in TMVA, features such as $\Delta \eta(j1, j2)$, $\Delta \phi(j1, j2)$, $\Delta R(j1, j2)$, $\Delta \eta(bj1, bj2)$, $\Delta \phi(bj1, bj2)$, and $\Delta R(bj1, bj2)$ were selected because they capture inter-jet correlations and angular separations, offering better cut-based discrimination in multivariate analysis (MVA) for signal-background separation.
\begin{table*}[htbp]
  \centering
  \resizebox{\textwidth}{!}{%
    \begin{tabular}{|c|c|}
      \hline
      \textbf{Variable} & \textbf{Description} \\
      \hline
      $N_j$ 
        & Number of light jets \\
      \hline
      $N_b$
        & Number of $b$-tagged jets \\
      \hline
      $H_T$ 
        & Scalar sum of $p_T$ of all final-state particles \\
      \hline
      $\makecell[l]{p_T(j_1),\,p_T(j_2), p_T(j_3),\,p_T(j_4) ,p_T(j_5) ,p_T(j_6)}$& Transverse momenta of the first six leading light jets\\
      \hline
      $p_T(bj_1)$ 
        & Transverse momentum of the leading $b$-jet \\
      \hline
      $\sum p_T(bj)$
        & Sum of $p_T$ of all $b$-tagged jets \\
      \hline
      $p_T^\text{miss}$ 
        & Missing transverse momentum \\
      \hline
      $\Delta \eta(j_1,j_2)$& Rapidity difference between the two leading light jets \\
      \hline
      $\Delta \phi(j_1,j_2)$ 
        & Azimuthal separation between the two leading light jets \\
      \hline
      $\Delta R(j_1,j_2)$
        & $\Delta R$ between the two leading light jets \\
      \hline
      $\Delta \eta(bj_1,bj_2)$& Rapidity difference between the two leading $b$-jets \\
      \hline
      $\Delta \phi(bj_1,bj_2)$ 
        & Azimuthal separation between the two leading $b$-jets \\
      \hline
      $\Delta R(bj_1,bj_2)$ 
        & $\Delta R$ between the two leading $b$-jets \\
      \hline
      $m_t$ 
        & Invariant mass of the three jets (including one $b$-tagged)
            closest to the top quark mass \\
      \hline
      $\displaystyle \frac{p_T^\text{miss}}{\sqrt{H_T}}$
        & Ratio of missing $p_T$ to $\sqrt{H_T}$ \\
      \hline
    \end{tabular}%
  }
  \caption{Key variables used in the analysis and their definitions\cite{saiel}.}
  \label{tab:dataset-variables}
\end{table*}
\twocolumngrid

For the proposed GNN architecture, the same features listed in table \ref{tab:dataset-variables} were utilized to ensure a consistent comparison with the BDT baseline. However, to construct the graph representation required for the GNN, individual $\eta$ and $\phi$ values of jets and b-jets were used to define nodes and edges, capturing spatial relationships between particles for the graph layers. In contrast, the BDT leveraged the derived features $\Delta \eta$, $\Delta \phi$, and $\Delta R$ for better cut-based performance. The BDT was trained in TMVA, while the GNN was implemented and trained in Python to accommodate its custom quantum-classical hybrid architecture and physics-informed Loss function.

\subsection{Data Preprocessing and Reweighting}

The preprocessing pipeline serves as the critical bridge between raw simulation outputs and robust deep learning analysis. We compute scale factors based on theoretical cross sections and integrated luminosity, ensuring that the subsequent classifier performance metrics (such as significance, calculated as $S/\sqrt{S+B}$) are normalized to reflect the real world data and is physically meaningful. 

The initial step in our data processing pipeline involved the extraction and conversion of the raw simulation outputs into a structured format suitable for subsequent machine learning analyses. The simulated events for both the signal process $t\bar{t}t\bar{t}W^-$ and its associated backgrounds were generated using a combination of state-of-the-art tools: MadGraph5 aMC@NLO was employed to calculate the matrix elements at leading (and next-to-leading) order, Pythia8 handled the parton showering and hadronization, and Delphes provided a fast yet realistic simulation of the CMS detector response. In the simulation output, each event was stored in a ROOT file and organized into separate trees corresponding to event classes based on signal and different backgrounds.

To bridge the gap between the high-energy physics simulation and the data requirements of modern deep learning frameworks, we utilized the UPROOT library to convert the ROOT file data into CSV format. This conversion process enabled the extraction of the raw feature information in a tabular form, thereby facilitating extensive exploratory data analysis and subsequent preprocessing steps in Python. The resultant CSV file comprises 21 columns in total as summarized by table \ref{tab:dataset-variables}. the 21st feature encodes the target variable, $event\_type\_encoded$, which categorizes each event into one of the eight distinct processes. 

The evaluation of classifier performance in high-energy physics is typically based on the metric of statistical significance, which is defined as 
\begin{equation}
    Z = \frac{S}{\sqrt{S+B}}
\end{equation}
where S and B represent the number of signal and background events, respectively. This metric provides a quantitative measure of how distinctly the signal can be separated from the background. The simulated dataset must be reweighted to reflect theoretical predictions. Table \ref{tab:cross_sections} presents the cross sections for each process, and in our study a luminosity of $350$ fb$^{-1}$ was considered. For instance, for the $t\bar{t}t\bar{t}W^-$ signal, the expected number of events is computed as:
\begin{equation}
    S_{expected} = \sigma_{signal} \times \mathcal{L}
\end{equation}
and similarly, the expected number of background events is given by the sum 
\begin{equation}
    B_{expected} = \sum_{i\neq signal}\sigma_i \times \mathcal{L}
\end{equation}
Since the simulated dataset contains a fixed number of events for each process, it is necessary to compute scale factors that normalize the event counts to these theoretical expectations. The scale factor for a given class i is determined by 
\begin{equation}
    SF_i = \frac{\sigma_i \times \mathcal{L}}{N_i}
\end{equation}
where $\sigma_i$ is the cross section for class i, $\mathcal{L}$ is the integrated luminosity, and $N_i$ is the total number of events of that class in the simulation. These scale factors are then applied when computing the significance so that the classifier performance reflects the realistic expected scenario. In our analysis, the signal $t\bar{t}t\bar{t}W^-$ corresponds to an expected yield of approximately 2.32 events, while the combined background yield is around 160 million events. 

\section{Physics-Informed Graph Neural Network for Multivariate Event Selection}\label{sec3}

The identification of rare signals like the four-top quark production with a W boson ($t\bar{t}t\bar{t}W^-$) amidst overwhelming Standard Model backgrounds is a significant challenge. This section introduces a quantum-enhanced, physics-informed graph neural network designed to address the limitations of multivariate classifiers like BDT and XGBoost by leveraging graph-based data representations, quantum machine learning, and physics-informed constraints. We begin with an overview of traditional cut-based classifiers as baselines, followed by a detailed description of the modified GNN architecture and its physics-informed loss functions.

\subsection{Overview of Cut-Based Classifiers}

Cut-based multivariate analysis remains a cornerstone in high-energy physics for event classification, leveraging predefined thresholds on key variables to isolate signals. This section examines two such methods: Boosted Decision Trees (BDTs) and XGBoost, optimized for the $t\bar{t}t\bar{t}W^-$ signal classification task. Both models were trained on a binary dataset encompassing the signal and background events, using features listed in Table \ref{tab:dataset-variables}.

The BDT implementation utilized the toolkit for multivariate analysis(TMVA) ROOT framework\cite{brun1997root}, with input features selected based on their physical relevance and discriminative power, as evaluated through their distributions in signal and background samples from Table \ref{tab:dataset-variables}, such as jet multiplicity, scalar $H_T$, and missing $E_T$. A grid search\cite{pedregosa2011scikit} further refined hyperparameters, yielding an optimal configuration of learning rate 0.1, maximum depth 3, 200 estimators, and a subsample ratio of 1.0. Similarly, the XGBoost\cite{Xgboost} model, implemented via the XGBClassifier from scikit-learn\cite{pedregosa2011scikit}, used the same set of input features and underwent grid search. These settings balanced model complexity and generalization, targeting effective signal-background separation.

While both models are robust and widely used in HEP, their reliance on predefined cuts and limited ability to model complex inter-particle correlations can constrain their performance for rare event classification. This motivates the development of advanced methods like GNN, which leverages graph-based representations  to capture intricate patterns in HEP data.

\subsection{Modified Physics-Informed Graph Neural Network}

Representing particle physics data as graphs emerges as a highly effective format, as it naturally encodes the spatial and kinematic relationships among particles, such as jets and b-jets, which are critical for distinguishing signal from background events. This subsection introduces the architecture of the modified GNN with classical graph layers and a quantum circuit. A quantum feature map embeds those features into the exponentially large Hilbert space of an n-qubit circuit to harness entanglement and interference effects. This quantum embedding can capture subtle, high order correlations among particle momenta and decay angles that a classical multilayer perceptron would require greatly increased depth or a much larger parameter count to approximate. In high energy physics analyses the available training data is often limited and full simulations are computationally costly. By using a shallow variational circuit with relatively few trainable gates we achieve a parameter efficient separation of signal and background. The result is enhanced discriminative power with reduced risk of overfitting compared to purely classical networks.

The GNN leverages a physics-informed Loss, namely the Jet Multiplicity Loss, which will be detailed in later subsections, to enforce physical consistency during training. By constructing graphs where nodes represent particles and edges encode their interactions, GNN employs a dual-stream mechanism that combines classical and quantum processing layers to enhance the discriminative power and interpretability of the model, as will be detailed in the following subsections.

\subsubsection{Representing Tabular Data as Graphs}

\begin{figure}
    \centering
    \includegraphics[width=1\linewidth]{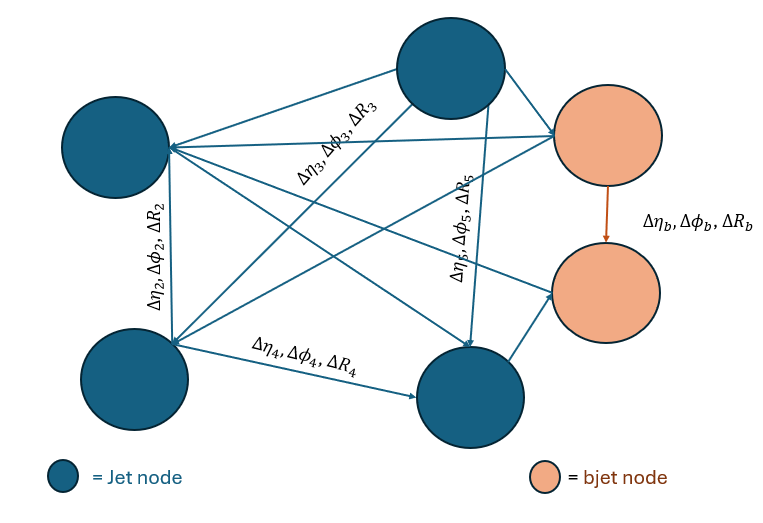}
    \caption{Schematic Diagram for data conversion from tabular to graph (An example of tabular to graph conversion).}
    \label{fig:tab-graph}
\end{figure}

Converting tabular particle physics data into a graph format is a crucial step before we begin the training for the GNN. Figure \ref{fig:tab-graph} illustrates this graph representation, where nodes are categorized into jet nodes (blue) and b-jet nodes (orange), with edges encoding their spatial relationships. Each event is transformed into a graph by treating particles as nodes, specifically jets and b-jets identified through features like \(P_T\), \(\eta\), and \(\phi\), with non-zero \(P_T\) values ensuring only physically present particles are included. For instance, the tabular data provides features for six jets and one b-jet, resulting in up to seven nodes per event, where each node’s feature vector comprises its \(P_T\), \(\eta\), and \(\phi\), chosen to encapsulate the kinematic properties essential for understanding particle trajectories and energy distributions. This choice reflects the physics of jet production, as higher \(P_T\) jets are more likely to originate from top quark decays in signal events, potentially enhancing the model’s ability to discern signal-specific patterns.

Edges between nodes are defined to form a fully connected graph, with edge features computed as \(\Delta \eta\), \(\Delta \phi\), and \(\Delta R = \sqrt{(\Delta \eta)^2 + (\Delta \phi)^2}\), where \(\Delta \eta\) and \(\Delta \phi\) are the differences in pseudorapidity and azimuthal angle between pairs of nodes, and \(\Delta \phi\) is adjusted to lie within \([-\pi, \pi]\) to account for the periodic nature of the azimuthal angle. This edge definition, as shown in Figure \ref{fig:tab-graph}, captures the angular separation between jets, which is physically significant because signal events like \(t\bar{t}t\bar{t}W^-\) often exhibit distinct jet clustering patterns (e.g., smaller \(\Delta R\) between jets from the same top quark decay) compared to backgrounds. Including \(\Delta R\) as an edge feature allows the GNN to learn these spatial relationships, potentially improving discrimination through better modeling of jet substructure. Additionally, global features such as numJets, numBJets, Scalar \(H_T\), and \(E_T^{miss}\) are attached to each graph, providing event-wide context that complements the node-level kinematics, ensuring the model can leverage both local particle interactions and global event properties during training. This graph construction sets a foundation for the GNN to exploit relational data, paving the way for its architectural design and implementation in the subsequent subsection.

\subsubsection{Architecture Design for the modified-GNN}

\begin{figure}
    \centering
    \includegraphics[width=1\linewidth]{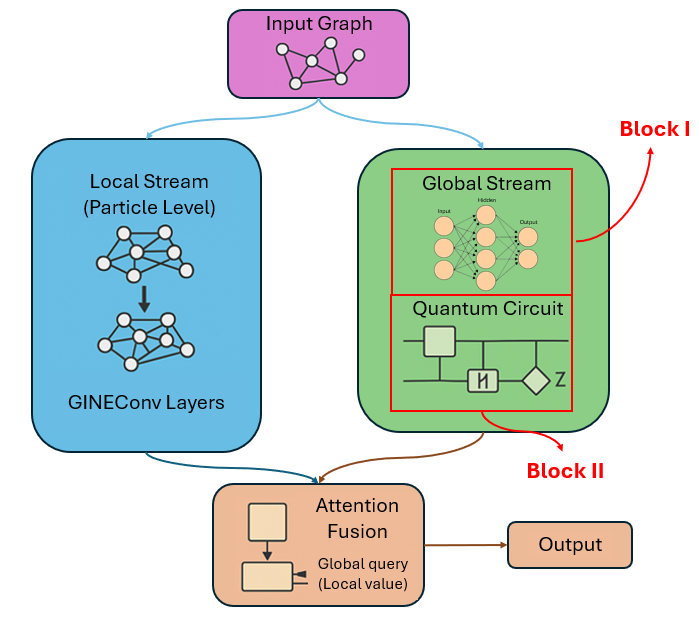}
    \caption{Schematic Diagram of the proposed architecture Modified GNN.}
    \label{fig:schem_CQPGNN}
\end{figure}

The architecture of the modified Graph Neural Network (GNN), depicted in Figure \ref{fig:schem_CQPGNN}, is designed to exploit the graph representation of \(t\bar{t}t\bar{t}W^-\) events by integrating classical graph neural network techniques with quantum computing principles, capturing both local particle interactions and global event properties while incorporating physics-informed constraints. The model processes an input graph where nodes represent jets and b-jets with features \(P_T\), \(\eta\), and \(\phi\), and edges encode spatial relationships via \(\Delta \eta\), \(\Delta \phi\), and \(\Delta R\), as established in the previous subsection. The architecture is divided into two primary streams: a local stream operating at the particle level using classical GNN layers, and a global stream handling event-wide features through both classical and quantum processing, both converging through a cross-attention mechanism to produce a classification output.

The local stream, as shown in Figure \ref{fig:schem_CQPGNN}, processes the input graph through two sequential GINEConv layers to extract node embeddings. The first GINEConv layer employs a multi-layer perceptron (MLP) with two linear layers (Linear \(\rightarrow\) ReLU \(\rightarrow\) Linear) to transform node features from the input dimension (\(node\_feature\_dim\)) to a hidden dimension (\(hidden\_dim = 64\)), while edge features (\(\Delta \eta\), \(\Delta \phi\), \(\Delta R\)) are processed by a parallel edge MLP of the same structure to match the hidden dimension. The output is passed through an Exponential Linear Unit (ELU)\cite{clevert2016fastaccuratedeepnetwork} activation and a dropout layer (with dropout probability \((p)=0.4\)) for regularization. The second GINEConv layer follows a similar structure, further refining the node embeddings with another MLP, ELU activation, and dropout. The node embeddings are then aggregated to the graph level by computing the mean across all nodes within each graph, resulting in a graph-level representation of dimension \(hidden\_dim\), which captures local particle phase space critical for identifying the jet-rich signature of \(t\bar{t}t\bar{t}W^-\) events.

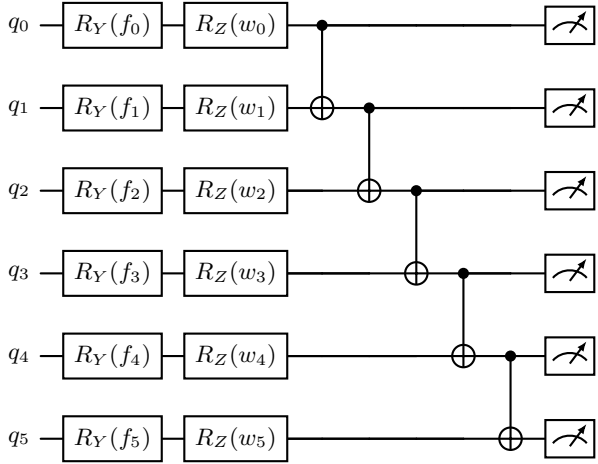
\begin{figure}[ht]
\centering
\begin{quantikz}[row sep={1.1cm,between origins}, column sep=0.3cm]
\lstick{$q_0$} & \gate{R_Y(f_0)} & \gate{R_Z(w_0)} & \ctrl{1} & \qw      & \qw      & \qw      & \qw      & \meter{} \\
\lstick{$q_1$} & \gate{R_Y(f_1)} & \gate{R_Z(w_1)} & \targ{}  & \ctrl{1} & \qw      & \qw      & \qw      & \meter{} \\
\lstick{$q_2$} & \gate{R_Y(f_2)} & \gate{R_Z(w_2)} & \qw      & \targ{}  & \ctrl{1} & \qw      & \qw      & \meter{} \\
\lstick{$q_3$} & \gate{R_Y(f_3)} & \gate{R_Z(w_3)} & \qw      & \qw      & \targ{}  & \ctrl{1} & \qw      & \meter{} \\
\lstick{$q_4$} & \gate{R_Y(f_4)} & \gate{R_Z(w_4)} & \qw      & \qw      & \qw      & \targ{}  & \ctrl{1} & \meter{} \\
\lstick{$q_5$} & \gate{R_Y(f_5)} & \gate{R_Z(w_5)} & \qw      & \qw      & \qw      & \qw      & \targ{}  & \meter{}
\end{quantikz}
\vspace{0.2cm}
\caption{Quantum encoder circuit: each classical feature $f_i$ is encoded via $R_Y(f_i)$ followed by a trainable $R_Z(w_i)$. Entanglement is introduced via nearest-neighbor CNOT gates to capture feature correlations. Outputs $\langle Z_i\rangle$ are measured and passed through a linear projection.}
\label{fig:Quantum_encoder}
\end{figure}

The global stream, shown in Figure \ref{fig:schem_CQPGNN}, processes event‑wide features in two parallel branches: a classical MLP and a quantum encoder. The classical branch applies two linear layers with a ReLU activation in between, mapping the input vector of length \(F\) into a \(D\)‑dimensional hidden embedding. The quantum branch is implemented in PennyLane\cite{bergholm2022pennylane} and uses exactly six qubits to match the six selected observables (\(\mathrm{numJets}\), \(\mathrm{numBJets}\), \(H_T\), $E_T^{miss}$, plus two summary features derived from node embeddings).  

Each observable \(f_i\) is first scaled into the interval \([-\pi,\pi]\). We then perform an \(R_Y\) rotation on qubit \(i\) to encode the real value into amplitude space:  
\begin{equation}
  R_Y(f_i)\ket{0} 
    = \cos\!\bigl(\tfrac{f_i}{2}\bigr)\ket{0}
    + \sin\!\bigl(\tfrac{f_i}{2}\bigr)\ket{1}.
  \label{eq:ry_rotation_revised}
\end{equation}  
This step is necessary because quantum states are manipulated via unitary rotations, and without it the raw kinematic value cannot influence the quantum amplitudes.  

Immediately after, a learnable phase shift \(R_Z(w_i)\) is applied on the same qubit:  
\begin{equation}
  R_Z(w_i) = \exp\!\bigl(-\tfrac{i\,w_i}{2}Z\bigr),
  \label{eq:rz_rotation_revised}
\end{equation}  
where \(w_i\) is a trainable parameter. These operations together form a single‑qubit layer for each index \(i=0,\dots,5\):  
\begin{equation}
  U_{\mathrm{single}}
    = \prod_{i=0}^{5} R_Z(w_i)\,R_Y(f_i).
  \label{eq:single_layer_revised}
\end{equation}  

To introduce correlations among observables, we entangle adjacent qubits with a chain of CNOT gates acting between qubit \(i\) and qubit \(i+1\) for \(i=0,\dots,4\):  
\begin{equation}
  U_{\mathrm{entangle}}
    = \prod_{i=0}^{4} \mathrm{CNOT}_{\,i,i+1}.
  \label{eq:entanglement_revised}
\end{equation}  
Using indices \(0\) through \(5\) for six qubits and entangling pairs \((i,i+1)\) ensures nearest‑neighbor correlations while keeping circuit depth minimal.  

The full quantum transformation is then  
\begin{equation}
  \ket{\psi}
    = U_{\mathrm{entangle}} \;U_{\mathrm{single}}\;\ket{0}^{\otimes6}.
  \label{eq:full_circuit_revised}
\end{equation}  
Finally, we measure each qubit in the computational basis to obtain expectation values  
\begin{equation}
  o_i = \langle Z_i\rangle 
      = \bra{\psi}Z_i\ket{\psi},
  \quad i=0,\dots,5,
  \label{eq:measurement_revised}
\end{equation}  
which form a six‑dimensional vector \(\mathbf{o}\in[-1,1]^6\). A subsequent linear projection  
\begin{equation}
  \mathbf{h}_q = W_q\,\mathbf{o} + b_q
  \label{eq:quantum_proj_revised}
\end{equation}  
then maps \(\mathbf{o}\) into the same \(D\)-dimensional hidden space as the classical branch.  

This circuit design balances expressive power and trainability. Six qubits suffice for one‑to‑one encoding of key observables. The \(R_Y\) rotations embed magnitudes into amplitudes. The \(R_Z\) gates introduce trainable phase flexibility. Nearest‑neighbor CNOTs capture the strongest pairwise correlations without excessive depth. Together, these choices yield a quantum encoding that complements the classical MLP and improves discrimination of the jet‑rich signatures in \(t\bar{t}t\bar{t}W^-\) events.  
 
The outputs from the local and global streams are fused using a cross-attention mechanism, as illustrated in Figure \ref{fig:schem_CQPGNN}. Here, \(G\in\mathbb{R}^{\text{hidden\_dim}}\) denotes the classical global embedding produced by the MLP branch, and \(Q\in\mathbb{R}^{\text{hidden\_dim}}\) denotes the quantum-enhanced global embedding from the quantum encoder. These serve as the query and key, respectively, while the graph-level node embedding \(L\in\mathbb{R}^{\text{hidden\_dim}}\) serves as the value. All three vectors are first normalized with layer normalization to stabilize training. The cross-attention output \(A\) is then computed as:
\begin{equation}
    A = \operatorname{softmax}\!\biggl(\frac{(W_Q\,G)(W_K\,Q)^{T}}{\sqrt{\text{hidden\_dim}}}\biggr)\,(W_V\,L),
    \label{eq:cross_attention}
\end{equation}
where \(W_Q\), \(W_K\), and \(W_V\) are learnable projection matrices, and we employ 8 attention heads to align the representations. This process, formalized in Equation \ref{eq:cross_attention}, allows the model to weigh the importance of local particle interactions (e.g., jet clustering patterns) against global event properties (e.g., \(MET\)), improving discrimination by focusing on features most relevant to \(t\bar{t}t\bar{t}W^-\) events, such as the presence of missing transverse energy from undetected particles. The attention output \(A\) is concatenated with the classical global embedding, resulting in a \(2 \times hidden\_dim\)-dimensional vector, which is then passed through a fusion fully connected (FC) layer. The fusion FC consists of two linear layers (Linear(128, 64) \(\rightarrow\) ReLU \(\rightarrow\) Dropout(0.4) \(\rightarrow\) Linear(64, 2)), producing a 2-class output. The final classification probabilities are obtained via a log-softmax activation:
\begin{align}
    p_c &= \text{log-softmax}\Bigl(W_2 \cdot 
    \text{ReLU}\bigl(W_1 \cdot [G; A] + b_1\bigr) + b_2\Bigr), \nonumber \\
    &\quad c \in \{0, 1\},
    \label{eq:log_softmax}
\end{align}

where \(W_1\), \(W_2\), \(b_1\), and \(b_2\) are the weights and biases of the fusion FC layers, and \(c=0\) represents the signal (\(t\bar{t}t\bar{t}W^-\)) and \(c=1\) the background. Equation \ref{eq:log_softmax} ensures the output is a normalized probability distribution suitable for binary classification, directly addressing the task of distinguishing the rare signal from dominant backgrounds. The training process incorporates Jet Multiplicity Loss, which will be detailed in later subsections, to ensure the model respects kinematic and jet multiplicity constraints, potentially improving classification performance by leveraging physical regularizations. This hybrid classical-quantum architecture, as shown in Figure \ref{fig:schem_CQPGNN}, offers a novel approach to modeling complex particle interactions in high-energy physics.

\subsection{Physics-Informed Losses}

This subsection introduces a physics-informed Loss function, Jet Multiplicity Loss which is designed to enhance the classification of \(t\bar{t}t\bar{t}W^-\) events by embedding physical constraints derived from the event decay dynamics into the training of the GNN. These losses regularize the model by enforcing consistency with expected signal versus background properties, and as we shall see later in the results section.

\subsubsection{Jet Multiplicity Loss}

The Jet Multiplicity Loss enforces consistency between the predicted jet multiplicity from the Monte Carlo (MC) simulations and expected jet multiplicities from the fully hadronic decay of \(t\bar{t}t\bar{t}W^-\) events compared to background processes. In the fully hadronic channel, the signal is characterized by a high jet multiplicity due to the decay of four top quarks and an additional W boson, producing approximately 4 b-jets and 10 light jets, while backgrounds like \(t\bar{t}\) yield fewer jets (e.g., 2 b-jets and 4 light jets), as summarized in Table \ref{tab:processes}. This loss guides the GNN to favor event reconstructions that match the expected jet counts. It improves the model’s focus on the jet-rich signal. Backgrounds with lower jet multiplicities are thus more effectively suppressed in the binary classification task.

\begin{table*}[htbp]
  \centering
  \resizebox{\textwidth}{!}{%
    \begin{tabular}{|c|c|c|c|c|}
      \hline
      \textbf{Process} 
        & \textbf{Decay Equations} 
        & \textbf{B jets}& \textbf{Light jets}& \textbf{Total jets}\\
      \hline
      $ttttW^{-}$ 
        & \makecell[l]{$4\,t \to 4\,(b\,W^{+}), \; W^{+}\to q\bar{q};$
                      $W^{-}\to q'\bar{q'}$}& 4 
        & 10 
        & 14 \\
      \hline
      $ttt\bar{b}$
        & \makecell[l]{$3\,t \to 3\,(b\,W^{+}), \; W^{+}\to q\bar{q};$\\
                      $\bar{b}$}
        & 4
        & 6
        & 10 \\
      \hline
      $tttW^{-}$
        & \makecell[l]{$3\,t \to 3\,(b\,W^{+}), \; W^{+}\to q\bar{q};$
                      $W^{-}\to q'\bar{q'}$}& 3
        & 8
        & 11 \\
      \hline
      $tttW^{+}$
        & \makecell[l]{$3\,t \to 3\,(b\,W^{+}), \; W^{+}\to q\bar{q};$
                      $W^{+}\to q'\bar{q'}$}& 3
        & 8
        & 11 \\
      \hline
      $t\bar{t}$
        & \makecell[l]{$t \to b\,W^{+}, \; \bar{t}\to \bar{b}\,W^{-},$
                      $W^{+}\to q\bar{q}, \; W^{-}\to q'\bar{q'}$}& 2
        & 4
        & 6 \\
      \hline
      $t\bar{t}H$
        & \makecell[l]{$t \to b\,W^{+}, \; \bar{t}\to \bar{b}\,W^{-},$
                      $W^{+}\to q\bar{q}, \; W^{-}\to q'\bar{q'};$
                      $H \to b\bar{b}$}& 4
        & 4
        & 8 \\
      \hline
      $t\bar{t}W^{-}$
        & \makecell[l]{$t \to b\,W^{+}, \; \bar{t}\to \bar{b}\,W^{-},$
                      $W^{+}\to q\bar{q}, \; W^{-}\to q'\bar{q'};$
                      $W^{-}\to q''\bar{q''}$}& 2
        & 6
        & 8 \\
      \hline
      $t\bar{t}Z$
        & \makecell[l]{$t \to b\,W^{+}, \; \bar{t}\to \bar{b}\,W^{-},$
                      $W^{+}\to q\bar{q}, \; W^{-}\to q'\bar{q'};$
                      $Z \to q''\bar{q''}$}& 2
        & 6
        & 8 \\
      \hline
    \end{tabular}%
  }
  \caption{Summary of the signal and background processes in hadronic decay modes and the expected number of \(b\)-jets, light jets, and total jets.}
  \label{tab:processes}
\end{table*}

Let \(N_j\) represent the number of light jets (numLightJets) and \(N_b\) the number of b-jets (numBJets), extracted from the global features. The expected counts (\(N^{exp}_l\), \(N^{exp}_b\)) and tolerances (\(tol_l\), \(tol_b\)) are defined based on the true label \(y_{true}\): for the signal (\(y_{true} = 0\)), \(N^{exp}_l = 10\), \(N^{exp}_b = 4\), \(tol_l = 1.668\), \(tol_b = 1.147\); for the background (\(y_{true} = 1\)), the maximum expected counts are \(N^{exp}_l = 8\), \(N^{exp}_b = 4\), with \(tol_l = 1.8\), \(tol_b = 1.2\). These tolerance values (\(tol_l\), \(tol_b\)) are set to one standard deviation of the jet‐count distributions observed in our simulated samples. Although the hadronic decay fixes the average light‑jet count at 10 and \(b\)‑jet count at 4 for the signal, Monte Carlo events exhibit natural fluctuations around these means. By using the 1$\sigma$ spread as our tolerance, the loss penalizes only those reconstructions that deviate beyond typical statistical variation, enforcing adherence to physically plausible jet multiplicities. The differences are computed as:

\begin{equation}
    diff_l = |N_j - N^{exp}_l|, \quad diff_b = |N_b - N^{exp}_b|,
\end{equation}

and the loss terms are calculated using a rectified squared error, with different conditions for signal and background to reflect their distinct jet profiles:

\begin{equation}
    L_{j} = \begin{cases} 
        (\text{ReLU}(diff_l - tol_l))^2 & \text{if } y_{true} = 0, \\
        (\text{ReLU}(N_j - E_l - tol_l))^2 & \text{if } y_{true} = 1,
    \end{cases}, \quad
\end{equation}
\begin{equation}
     L_{b} = \begin{cases} 
        (\text{ReLU}(diff_b - tol_b))^2 & \text{if } y_{true} = 0, \\
        (\text{ReLU}(N_b - E_b - tol_b))^2 & \text{if } y_{true} = 1,
    \end{cases},
\end{equation}

where \(\text{ReLU} = \max(0, x)\). The total loss is weighted by the model’s predicted probability \(p_{pred}\) for the predicted class and averaged over the batch:

\begin{equation}
    L_{jet\_mult} = \frac{1}{B} \sum_{i=1}^B p_{pred,i} \cdot (loss_{light,i} + loss_{b,i}),
\end{equation}

where \(B\) is the batch size. By penalizing deviations from expected jet counts, this loss ensures the GNN learns to recognize the high jet multiplicity of \(t\bar{t}t\bar{t}W^-\) events, improving its discriminative power in the MVA by focusing on a key physical signature of the signal.

\section{Results and Discussion}

This section evaluates the performance of the GNN in classifying \(t\bar{t}t\bar{t}W^-\) events, focusing on its training procedure, classification results, and output distributions, followed by a comparison with a baseline Boosted Decision Tree model to highlight its effectiveness.

\subsection{Training Setup for GNN}

\begin{figure}
    \centering
    \includegraphics[width=1\linewidth]{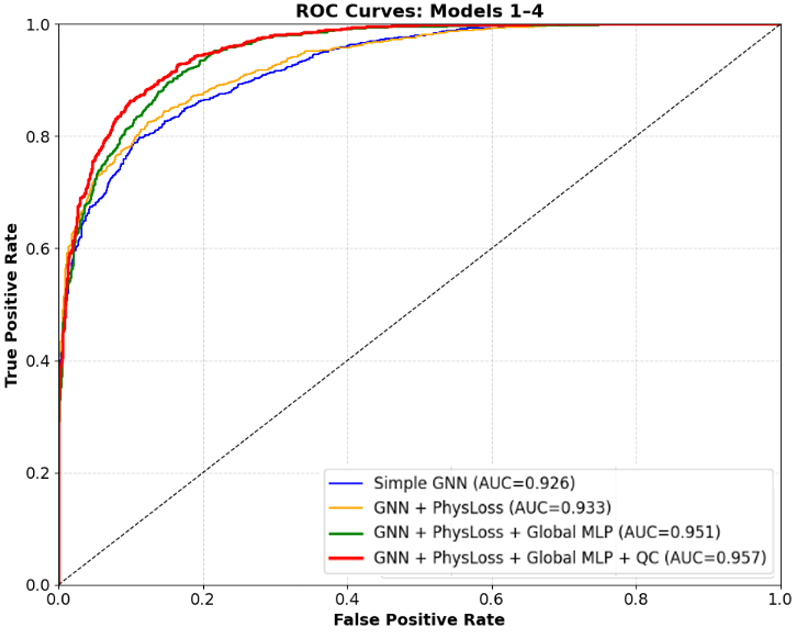}
    \caption{The ROC AUC curves comparison of the GNN model variants, depicting the improvements made by additional components.}
    \label{fig:roc_variants}
\end{figure}

A preliminary study was carried out. It compared four model variants under identical training conditions to isolate the impact of each architectural component. The first variant used only two GINEConv layers. The second variant augmented this backbone with physics informed jet multiplicity Loss. The third variant further incorporated a classical global feature multilayer perceptron corresponding to Block I. The fourth variant completed the final modified GNN by adding the quantum circuit of Block II. Each model was trained for ten epochs and their ROC curves are shown in Figure~\ref{fig:roc_variants}. Table~\ref{tab:variant_comparison} summarizes their recall and ROC AUC scores.

\begin{table}[htbp]
  \centering
  \caption{Preliminary comparison of model variants over 10 epochs.}
  \label{tab:variant_comparison}
  \begin{tabular}{lcc}
    \hline
    \textbf{Model Variant}          & \textbf{Recall} & \textbf{ROC AUC} \\
    \hline
    GNN (Base)                      & 0.83            & 0.926            \\
    GNN (with PhysLoss)             & 0.89            & 0.933            \\
    GNN (Block I)                   & 0.91            & 0.951            \\
    GNN (Block I + II)              & 0.95            & 0.957            \\
    \hline
  \end{tabular}
\end{table}

As shown in Table~\ref{tab:variant_comparison} and Figure~\ref{fig:roc_variants}, adding physics informed losses increases recall from 0.83 to 0.89 and raises ROC AUC from 0.926 to 0.933. Including the global feature multilayer perceptron in Block I further boosts recall to 0.91 and lifts ROC AUC to 0.951. Finally, integrating the quantum circuit in Block II delivers the highest performance with recall of 0.95 and ROC AUC of 0.957. These stepwise gains demonstrate the value of each component in enhancing classification of the rare four top signal.\\
\\
The GNN was trained using a combination of standard and physics informed loss functions to optimize its performance for binary classification of signal versus background events. The final optimization of the modified GNN process employed the Adam\cite{kingma2017adam} optimizer with an initial learning rate of \(5 \times 10^{-4}\), paired with a cosine annealing learning rate scheduler over a maximum of 20 epochs (\(T_{max} = 20\)). The primary classification Loss was the Negative Log-Likelihood (NLL) Loss, augmented by the Jet Multiplicity Loss, with their contributions dynamically weighted based on the model’s precision and recall. Specifically, the weights for the physics-informed Loss were adjusted using target precision and recall values of 0.85 and 0.95, respectively, resulting in the jet multiplicity Loss weights from 0.05 to 0.3. This dynamic weighting, guided by an exponential moving average (EMA) of precision and recall with a decay of 0.9, ensured a focus on minimizing false positives, crucial for rare event detection. Additionally, an EMA with a decay of 0.99 was applied to the model parameters, stabilizing training by maintaining a smoothed version of the weights, particularly beneficial for the quantum circuit’s parameters, which were initially frozen and enabled for gradient updates after the third epoch. Training was conducted on a Kaggle Python notebook with no accelerator enabled, ensuring efficient computation over 20 epochs, with validation accuracy monitored to assess convergence.

\subsection{Performance and Output Analysis of GNN}

\begin{figure}[htbp]
    \centering
    \includegraphics[width=1\linewidth]{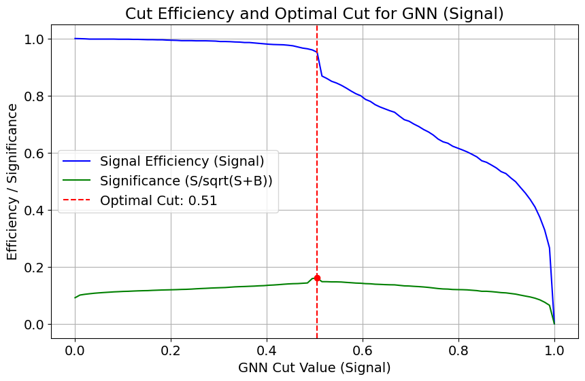}
    \caption{Cut efficiency and optimal cut for GNN, showing signal efficiency (\(S/(S+B)\)) and significance (\(S/\sqrt{S+B}\)) as functions of the GNN cut value, with an optimal cut at 0.51.}
    \label{fig:gnn_significance}
\end{figure}

The GNN achieved strong performance in classifying \(t\bar{t}t\bar{t}W^-\) events, with a significance of 0.174, a recall of 0.957, and a ROC-AUC score of 0.974, demonstrating its effectiveness in identifying the rare signal while maintaining high discriminative power. The significance, defined as \(S/\sqrt{S+B}\), where \(S\) and \(B\) are the signal and background yields, respectively, was optimized by determining an optimal cut on the GNN output. Figure \ref{fig:gnn_significance} illustrates the signal efficiency (\(S/(S+B)\)) and significance as functions of the GNN cut value, identifying an optimal cut at 0.51, which balances signal retention with background suppression, maximizing the model’s ability to detect \(t\bar{t}t\bar{t}W^-\) events.

\begin{figure}[htbp]
    \centering
    \includegraphics[width=1\linewidth]{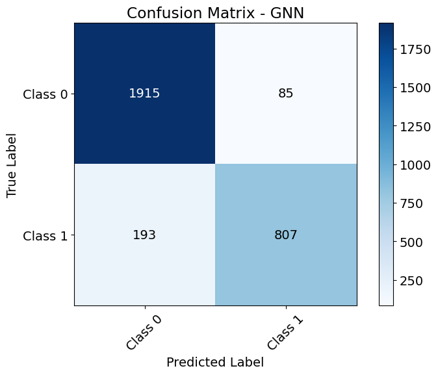}
    \caption{Confusion matrix for 
GNN, showing true labels (Class 0: Signal, Class 1: Background) versus predicted labels.}
    \label{fig:gnn_confusion}
\end{figure}

The confusion matrix in Figure \ref{fig:gnn_confusion} further details the classification performance, with true positives (TP) of 1915, false negatives (FN) of 85, false positives (FP) of 193, and true negatives (TN) of 807. This yields a precision of \(TP/(TP+FP) = 1915/(1915+193) \approx 0.908\) and an accuracy of \((TP+TN)/(TP+TN+FP+FN) = (1915+807)/(1915+807+193+85) \approx 0.906\). The high recall of 0.957 (\(TP/(TP+FN) = 1915/(1915+85)\)) underscores the model’s ability to capture most signal events, critical for rare event searches, while the low false negative rate ensures minimal signal loss.

\begin{figure}[htbp]
    \centering
    \includegraphics[width=1\linewidth]{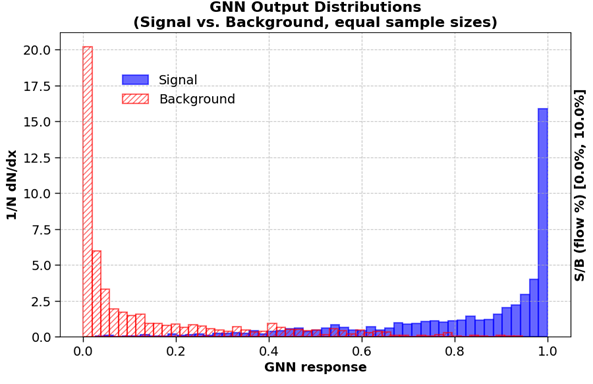}
    \caption{GNN output distributions for signal and background events (equal sample sizes), demonstrating clear separation with signal peaking near 1.0 and background near 0.}
    \label{fig:gnn_output_dist}
\end{figure}

The output distribution of the GNN, shown in Figure \ref{fig:gnn_output_dist}, reveals clear separation between signal and background events. Background events peak near 0, while signal events peak near 1.0, indicating that the model effectively assigns high probabilities to the correct classes, facilitating robust discrimination in the fully hadronic channel.

\subsection{Comparison with Baseline BDT}

\begin{figure}
    \centering
    \includegraphics[width=1\linewidth]{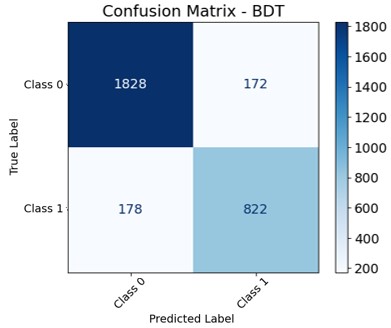}
    \caption{Confusion matrix for 
BDT, showing true labels (Class 0: Signal, Class 1: Background) versus predicted labels.}
    \label{fig:BDT_confusion}
\end{figure}

To contextualize the GNN’s performance, we compare it with a baseline Boosted Decision Tree (BDT), a standard method in high-energy physics as a part of TMVA, and XGBoost, another gradient-boosting approach, both trained on the same dataset with default hyperparameters. The GNN outperforms both baselines across key metrics: it achieves a significance of 0.174 compared to 0.148 for the BDT and 0.149 for XGBoost, a recall of 0.957 versus 0.914 (BDT) and 0.924 (XGBoost), and a ROC-AUC score of 0.96 against 0.91 (BDT) and 0.92 (XGBoost). This represents a 17.6\% improvement in significance over the BDT (\((0.174-0.148)/0.148\)) and a 4.7\% improvement in recall, highlighting the GNN’s superior signal identification and discriminative power.

The BDT’s confusion matrix in Figure \ref{fig:BDT_confusion}, shows true positives of 1828, false negatives of 172, false positives of 178, and true negatives of 822, resulting in a lower recall (\(1828/(1828+172) = 0.914\)) and accuracy (\((1828+822)/(1828+822+178+172) \approx 0.883\)) compared to the GNN. The higher false negative rate (172 vs. 85) indicates that the BDT misses more signal events, which is detrimental for rare event detection.

\begin{figure}[htbp]
    \centering
    \includegraphics[width=1\linewidth]{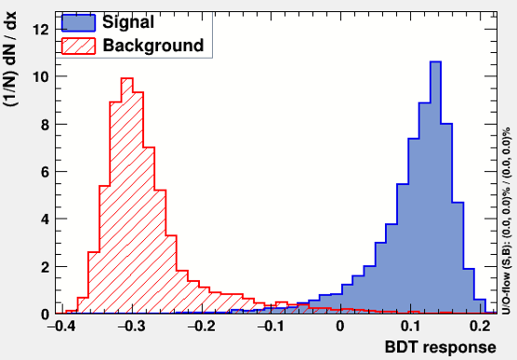}
    \caption{BDT output distributions for signal and background events, showing more overlap compared to the GNN, with signal peaking near -0.3 and background near 0.1\cite{saiel}.}
    \label{fig:bdt_output_dist}
\end{figure}

The output distributions further illustrate the GNN’s advantage. Figure \ref{fig:bdt_output_dist} shows the BDT’s response, with signal peaking near -0.3 and background near 0.1, exhibiting more overlap than the GNN’s distribution (Figure \ref{fig:gnn_output_dist}). This greater overlap suggests poorer separation, limiting the BDT’s ability to distinguish \(t\bar{t}t\bar{t}W^-\) events from background.

\begin{figure}[htbp]
    \centering
    \includegraphics[width=1\linewidth]{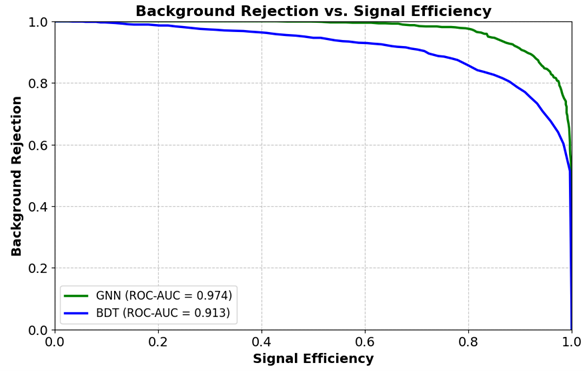}
    \caption{Background rejection versus signal efficiency for GNN (AUC = 0.974) and BDT (AUC = 0.913), showing the GNN’s superior discriminative performance.}
    \label{fig:roc_comparison}
\end{figure}

The ROC curve in Figure \ref{fig:roc_comparison} compares the background rejection versus signal efficiency for the GNN and BDT. The GNN achieves a higher AUC compared to the BDT, demonstrating a better trade-off between signal retention and background suppression, which is crucial for enhancing the detection of rare events like \(t\bar{t}t\bar{t}W^-\). 

Table \ref{tab:results_summary} summarizes the performance metrics, confirming the GNN’s superiority across all evaluated metrics. The hybrid classical-quantum approach, combined with physics-informed Loss, enables the GNN to better capture the complex patterns of \(t\bar{t}t\bar{t}W^-\) events, making it a more effective tool for rare event classification in high-energy physics.

\begin{table}[htbp]
\centering
\begin{tabular}{|l|c|c|c|}
\hline
\textbf{Model} & \textbf{Significance} & \textbf{Recall} & \textbf{ROC-AUC Score} \\
\hline
BDT& 0.148 & 0.914 & 0.913\\
XGBoost & 0.149 & 0.924 & 0.920\\
GNN& 0.174 & 0.957 & 0.974\\
\hline
\end{tabular}
\caption{Summary of performance metrics for GNN, BDT, and XGBoost, highlighting the GNN’s superior performance in significance, recall, and ROC-AUC score.}
\label{tab:results_summary}
\end{table}

\begin{figure}
    \centering
    \includegraphics[width=1\linewidth]{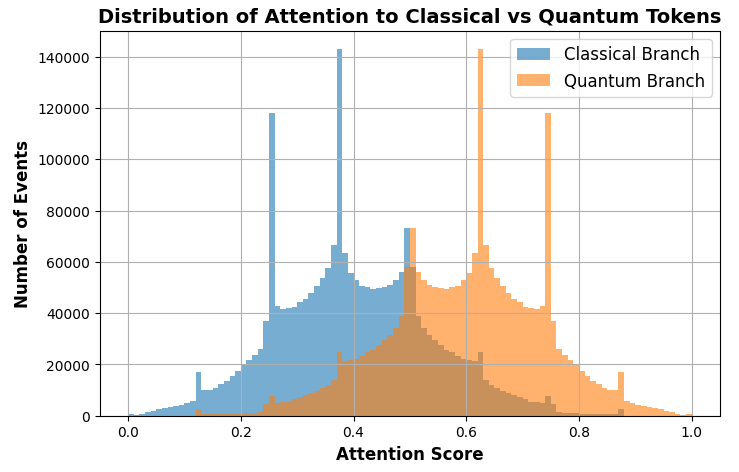}
    \caption{Histogram showing the distribution of attention scores assigned to the classical (blue) and quantum (orange) tokens across all events.}
    \label{fig:attention_distribution}
\end{figure}
To understand which information the model relies on more, we looked at how much attention it gives to the classical and quantum branches. As shown in Figure~\ref{fig:attention_distribution}, the model tends to focus more on the quantum token than the classical one. On average, the quantum token receives about 60\% of the attention, while the classical token gets around 40\%. This means that, during decision-making, the model often finds the quantum information more useful. Such analysis helps us see that the quantum branch of the model is not only being used, but also plays an important role in making predictions.

\section{Conclusion}
We have presented a novel, physics-informed, classical–quantum hybrid Graph Neural Network for the challenging task of fully hadronic \(t\bar{t}t\bar{t}W^-\) event classification at $\sqrt{s} = 13$ TeV. By combining GINEConv layers to capture local jet correlations, a six-qubit quantum circuit to encode global event observables via parameterized rotations and entanglement, and a cross-attention fusion mechanism, our model learns a richly structured representation that respects expected jet multiplicities. When benchmarked with 350 $fb^{-1}$ of MC-simulated data, it delivers a signal significance of (0.174), recall of (0.957) and ROC-AUC of (0.974), outperforming both BDT (0.148, 0.914, 0.913) and XGBoost (0.149, 0.924, 0.913). These results demonstrate that embedding physical priors and quantum-enhanced global feature processing within a GNN framework can substantially boost sensitivity to ultra-rare Standard Model processes, opening a promising avenue for precision measurements and new physics searches at the LHC.

\FloatBarrier

\bibliography{four_top}
\FloatBarrier

\end{document}